\documentclass[fleqn,usenatbib]{mnras}

\usepackage{newtxtext,newtxmath}

\usepackage[T1]{fontenc}

\DeclareRobustCommand{\VAN}[3]{#2}
\let\VANthebibliography\thebibliography
\def\thebibliography{\DeclareRobustCommand{\VAN}[3]{##3}\VANthebibliography}

\usepackage{graphicx}	
\usepackage{amsmath}	

\usepackage{dcolumn}
\usepackage{bm}
\usepackage{hyperref}
\usepackage{siunitx}
\usepackage{color}
\usepackage{commath}
\usepackage{xcolor}
\usepackage{soul}
\usepackage{natbib}
\usepackage{booktabs}

\title[Imaging fermionic dark matter cores at the center of galaxies]{Imaging fermionic dark matter cores at the center of galaxies}

\author[J. Pelle et al.]{
J. Pelle,$^{1,2}$\thanks{E-mail: jpelle@mi.unc.edu.ar}
C. R. Argüelles,$^{3,4}$\thanks{E-mail: carguelles@fcaglp.unlp.edu.ar}
F. L. Vieyro,$^{5,6}$
V. Crespi,$^{3,6}$
C. Millauro,$^{7}$
M. F. Mestre,$^{3,6}$
O. Reula,$^{1,2}$
\newauthor{and F. Carrasco$^{1,2}$}
\\
$^{1}$Facultad de Matemática, Astronomía, Física y Computación, Universidad Nacional de Córdoba, Argentina\\
$^{2}$Instituto de Física Enrique Gaviola, CONICET, Ciudad Universitaria, 5000 Córdoba, Argentina\\
$^{3}$Instituto de Astrofísica de La Plata, UNLP \& CONICET, Paseo del Bosque, B1900FWA La Plata, Argentina\\
$^{4}$ICRANet, Piazza della Repubblica 10, 65122 Pescara, Italy\\
$^{5}$Instituto Argentino de Radioastronomía (IAR, CONICET/CIC/UNLP), C.C.5, (1894) Villa Elisa, Buenos Aires, Argentina\\
$^{6}$Fac. de Ciencias Astron. y Geofísicas, Universidad Nacional de La Plata, Paseo del Bosque, B1900FWA La Plata, Argentina\\
$^{7}$Departamento de Física, Facultad de Ciencias Exactas y Naturales, Universidad de Buenos Aires, Pabellón I, Ciudad Universitaria, 1428 Buenos Aires, Argentina
}

\date{Accepted XXX. Received YYY; in original form ZZZ}

\pubyear{\the\year{}}

\begin{document}
\label{firstpage}
\pagerange{\pageref{firstpage}--\pageref{lastpage}}
\maketitle

\begin{abstract}
Current images of the supermassive black hole (SMBH) candidates at the center of our Galaxy and M87 have opened an unprecedented era for studying strong gravity and the nature of relativistic sources. Very-long-baseline interferometry (VLBI) data show images consistent with a central SMBH within General Relativity (GR). However, it is essential to consider whether other well-motivated dark compact objects within GR could produce similar images. Recent studies have shown that dark matter (DM) halos modeled as self-gravitating systems of neutral fermions can harbor very dense fermionic cores at their centers, which can mimic the spacetime features of a black hole (BH). Such dense, horizonless DM cores can satisfy the observational constraints: they can be supermassive and compact and lack a hard surface. We investigate whether such cores can produce similar observational signatures to those of BHs when illuminated by an accretion disk. We compute images and spectra of the fermion cores with a general-relativistic ray tracing technique, assuming the radiation originates from standard $\alpha$ disks, which are self-consistently solved within the current DM framework. Our simulated images possess a central brightness depression surrounded by a ring-like feature, resembling what is expected in the BH scenario. For Milky Way-like halos, the central brightness depressions have diameters down to $\sim 35\, \mu\text{as}$ as measured from a distance of approximately $8\,$kpc. Finally, we show that the DM cores do not possess photon rings, a key difference from the BH paradigm, which could help discriminate between the models.
\end{abstract}




\section{Introduction}
\label{sec:Intro}

Three different observational campaigns aimed at the galaxy center have confirmed that Sgr A* must be a supermassive compact object of roughly $4\times 10^6 M_\odot$. Two independent campaigns met this conclusion through the study of stellar motions around Sgr A* \citep{2005ApJ...620..744G, 2008ApJ...689.1044G, 2010RvMP...82.3121G,2018A&A...615L..15G,2019Sci...365..664D,2020A&A...636L...5G}. More recently, a third and independent campaign corroborated the same mass inference for Sgr A* by observing the relativistic images caused by lensed photons on event-horizon scales via VLBI techniques \citep{akiyama2022first,akiyama2022-IIIfirst,akiyama2022-IVfirst}. These results add up to the first image of the supermassive black hole (SMBH) candidate at the center of the giant galaxy M87 \citep{2019ApJ...875L...1E} which, unlike Sgr A*, is the source of an active galactic nucleus. Even though the spacetime geometry associated with the two objects mentioned above is consistent with that generated by a Kerr black hole (BH) \citep{2019ApJ...875L...1E,2022ApJ...930L..12E, boero2021strong}, considerable efforts have been made to provide alternative candidates which may explain the observational signatures \citetext{see \citealp{cardoso2019} for a review}. These alternatives include the so-called \textit{gravastars} \citep{mazur2004,visser2004,cattoen2005} motivated within quantum gravity theories; boson stars \citep{vincent2016,2020MNRAS.497..521O,vincent2021,rosa2023imaging}; and compact objects made of dark matter (DM), 
either self-interacting \citep{2016MNRAS.461.4295S,2016JCAP...04..038A,2020PDU....3000699Y} or self-gravitating systems of semi-degenerate neutral fermions \citep{arguelles2018,2019PDU....24..278A,2020A&A...641A..34B,2021MNRAS.505L..64B,2021MNRAS.502.4227A,2022MNRAS.511L..35A}, among others.

In this work, we center our attention on the latter case. We investigate if high-density concentrations of DM fermions, which naturally arise at halo centers, provide gravitational light-bending signatures similar to that of a BH. The motivations for this choice are both theoretical and observational. On the theoretical side, the fermionic DM halo model as defined in \citet{2015MNRAS.451..622R,arguelles2018} incorporates the quantum (fermionic) nature of the particles (not feasible in standard N-body simulations). It self-consistently considers the Pauli exclusion principle, thus yielding a quantum pressure that dominates over the central region of the configurations. As a result, the model predicts novel DM density profiles with a \textit{dense core}--\textit{diluted halo} morphology that depends on the fermion mass. On the observational side, distributions with fermion masses in the $\sim 50-350\,$keV range which naturally account for the large-scale structure of the Universe, can also explain the galaxy rotation curves in different types of galaxies  \citep{arguelles2018,2019PDU....24..278A,2023ApJ...945....1K}. At the same time, the degenerate fermion core residing at the halo center can mimic their central BHs \citep{arguelles2018,2019PDU....24..278A,2020A&A...641A..34B,2021MNRAS.505L..64B,2021MNRAS.502.4227A,2022MNRAS.511L..35A,2022IJMPD..3130002A}, or eventually collapse into one as demonstrated in \citet{CHAVANIS2020135155,2021MNRAS.502.4227A,2023MNRAS.523.2209A,2024ApJ...961L..10A} from general-relativistic stability criteria. 

In the context of active galaxies, \citet{millauro2024} extends the thin disk solutions of Shakura \& Sunyaev \citep{shakura1973} to the case of the above-mentioned core--halo fermionic DM model. This work demonstrates two key results within the fermionic scenario: (i) For a given DM core mass, there exists certain core compactness---i.e., corresponding fermion mass---which produces a luminosity spectrum that is essentially indistinguishable from that of a Schwarzschild BH of the same mass; and (ii) the disk can enter the nonrotating core, achieving accretion efficiencies as high as $28\%$, comparable to those of rapidly rotating Kerr BHs. 

The next step, which is the subject of the present paper, is to study the relativistic images produced by the lensed photons emerging from this extended thin disk solution around fermionic cores. With these two general results---the luminosity spectra of $\alpha$--disks and the strong-field images and luminosity patterns around the fermionic cores---we complement the analogous sequence previously shown for boson stars in \citet{2006PhRvD..73b1501G,vincent2016}, though in our case applied to typical active-like galaxies.

In this work, we adopt the geometrized unit system where the gravitational constant and the vacuum speed of light are set to unity ($G=c=1$). 

The article is organized as follows: in Sect.~\ref{sec:model}, we briefly describe the fermionic halo model and the accretion disk solutions around the compact fermion cores. In Sect.~\ref{sec:methods}, we present our numerical methods for producing the images and spectra based on a general-relativistic ray tracing technique adapted to the spacetime of our interest. In Sect.~\ref{sec:parameters}, we describe the parameters of the various configurations we explore. In Sect.~\ref{sec:results}, we show the results of the different relativistic images cast by the fermion cores for various viewing angles and fermion masses. In Sect.~\ref{sec:discussion}, discuss the results. Finally, in Sect.~\ref{sec:conclusions}, we give some conclusions and outline future perspectives.

\section{Physical model}\label{sec:model}

\subsection{Fermionic dark matter model for galactic halos}
\label{sec:RAR}

The fermionic DM halo model under consideration here is commonly referred to as the Ruffini--Argüelles--Rueda (RAR) model \citep{2015MNRAS.451..622R}. We apply its extended version, which accounts for the effects of particle escape \citep{arguelles2018}. 
This model is sometimes called the relativistic fermionic--King model \citep{2022PhRvD.106d3538C}, and it describes a self-gravitating system of massive, neutral fermions (spin 1/2) in hydrostatic equilibrium, incorporating particle escape within the framework of GR. 
These fermions follow a distribution function of the Fermi-Dirac type with a cut-off given by
\begin{equation}
    f_c(\epsilon \leq \epsilon_c,r) = \frac{1- e^{(\epsilon - \epsilon_c)/k_B T(r)}}{e^{(\epsilon - \mu)/k_B T(r)}+1}\,,  \quad f_c(\epsilon > \epsilon_c , r) = 0\,,
\label{eq:DF}
\end{equation}
where $\epsilon = \sqrt{p^2 + m^2} - m$ is the particle kinetic energy, $\epsilon_c$ is the cut-off kinetic energy, $\mu$ is the chemical potential with the rest energy subtracted off, $T(r)$ is the temperature, and $c$, $k_B$, $m$ are the speed of light, the Boltzmann constant, and the fermion mass respectively. 
This distribution can be derived from a maximum entropy production principle as first shown in \citet{chavanis1998}, and recently successfully applied to a large set of disk-galaxies rotation curves in \citet{2023ApJ...945....1K}. 

\begin{figure}
    \centering
    \includegraphics[width = \columnwidth]{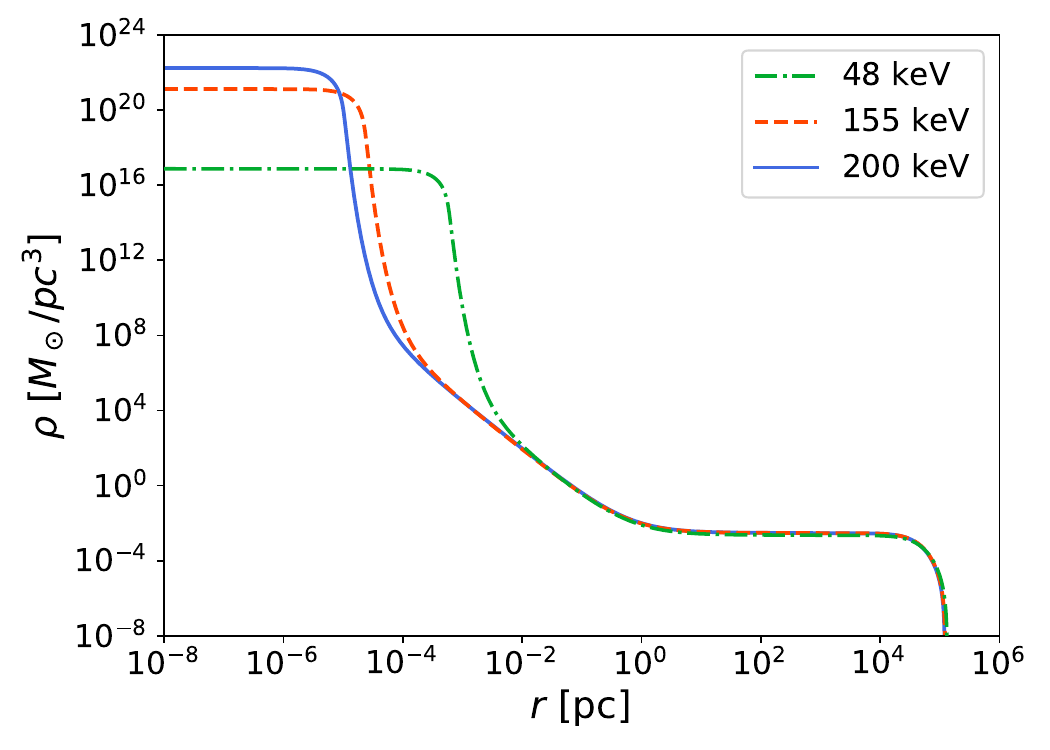}
    \caption{Density core-halo profiles for fermion masses of 48 keV (dotted-dashed green), 155 keV (dashed orange), and 200 keV (solid blue). All solutions correspond to a core mass of $M_{\rm c}=1\times 10^7 M_\odot$ and a total mass of $M_T=1\times 10^{12} M_\odot$ at $r_T=122\,$kpc, consistent with the Ferrarese relation between the dark central object and the DM halo \citep{ferrarese2002}.}
    \label{fig:rho}
\end{figure}

In this model, the system is regarded as static and spherically symmetric, so the spacetime metric is parameterized as
\begin{equation}
    ds^2 = -e^\nu dt^2 + e^\lambda dr^2 + r^2 (d\theta^2 + \sin^2\!\theta d\phi^2)\,,
    \label{eq:rar_metric}
\end{equation}
where $\nu(r)$ and $\lambda(r)$ are functions to be solved along with the hydrostatic equilibrium equations for the fermionic halo. These equations include the Tolman--Oppenheimer--Volkoff (TOV) equations for a perfect fluid, combined with the Tolman and Klein conditions (representing the zeroth and first laws of thermodynamics in GR) and energy conservation along geodesics. The resulting system of equations \citep[see Sect. 2 of][]{arguelles2018} involves a numerical boundary condition problem,
for a given DM particle mass\footnote{See also Appendix A in \citet{2024arXiv240419102M} for details on numerically solving these equations.}. There exists a family of RAR profiles that develop a \textit{dense core}--\textit{diluted halo} morphology, where fermion-degeneracy pressure supports the core and thermal pressure supports the halo \citep{arguelles2018,2019PDU....24..278A,2021MNRAS.502.4227A}. We adopt the particle mass range from approximately $50$ to $350\,$keV because, as mentioned in Sect.~\ref{sec:Intro}, the dense fermion cores within this range can mimic the BH candidate in Sgr A* while the outer halos explain the Galaxy rotation curve \citep{arguelles2018,2019IJMPD..2843003A}. Previous works have also used particle masses within this range to model other types of systems ranging from dwarf galaxies to galaxy clusters \citetext{see, e.g., \citealp{2019PDU....24..278A,2023MNRAS.523.2209A,2023ApJ...945....1K} and \citealp{2023Univ....9..197A} for a review}. 
Moreover, larger fermion masses imply more compact cores for a fixed halo boundary condition and a fixed core mass $M_{\rm c}$. We show this in Fig. \ref{fig:rho} for the cases $m=48\,$keV, $m=155\,$keV, and $m=200\,$keV, with $M_{\rm c}= 10^7 M_\odot$. It is worth mentioning that these solutions, having a total halo mass of $10^{12} M_\odot$, are in agreement with the observational relation between halo mass and the masses of their corresponding supermassive central objects \citep{ferrarese2002}. 

Since the RAR model is built upon a self-gravitating system of tempered fermions under a perfect fluid ansatz within GR, sufficiently degenerate fermion cores can reach a critical mass of core-collapse into a SMBH \citep{2021MNRAS.502.4227A,2023MNRAS.523.2209A,2024ApJ...961L..10A}. This critical core mass is $M_{\rm c}^{cr}\approx M_{\mathrm{OV}}=0.384 m_{\mathrm{pl}}^3/m^2$, where $M_{\mathrm{OV}}$ is the Oppenheimer-Volkoff mass and $m_{\mathrm{pl}}$ is the Planck mass, as shown in the above references with applications to the problem of SMBH formation in the early Universe. In the specific case of interest for active galaxies with a central object mass of $M_{\rm c}= 10^7 M_\odot$, a particle mass of $m=200\,$keV implies a compact DM core that is very close to its critical value for collapse.       

\subsection{The thin accretion disk}
\label{sec:accretion-disk}

The solution obtained in \citet{millauro2024} is the extension of the standard model of a thin disk solution \citep{shakura1973} to the context of regular (i.e., not singular) compact mass distributions. In this solution, the orbital angular velocity is given by the Newtonian expression\footnote{The relativistic correction to the angular velocity is around $0.1 \%$ or below in all scenarios considered in this work.} 
\begin{equation}
    \Omega = \left(\frac{G M(r)}{r^3} \right)^{1/2}\,,
    \label{eq:angular_speed}
\end{equation}
\noindent and a small (highly subsonic) component of radial velocity. The disk is thin, with a height scale $H \ll r$, and the viscosity is parameterized with the $\alpha$-prescription, where the kinematic viscosity $\nu$ is given by $\nu = \alpha c_{\rm s} H$. Here, $c_{\rm s}$ is the speed of sound, and $\alpha$ is a dimensionless parameter smaller than one.

Under these assumptions, the disk is optically thick in the vertical direction, and each disk element is assumed to radiate as a blackbody at a radius-dependent temperature $T(r)$. The temperature profile around the DM core solution is given by

\begin{equation}\label{eq:temperature}
\begin{aligned}
    T(r) = \Biggl\{ \frac{3\Dot{M}}{8\pi \sigma}\frac{GM(r)}{r^3}
    & \left[1- \left(\frac{M_{\text{in}}r_{\text{in}}}{M(r)r}\right)^{1/2}\right] \\
    \times & \left[1-\frac{r}{3M(r)} \frac{dM(r)}{dr}\right] \Biggl\}^{1/4}\,,
\end{aligned}
\end{equation}

\noindent where $\dot{M}$ is the accretion rate, $r_{\text{in}}$ is the inner radius of the disk, and $\sigma$ is the Stefan-Boltzmann constant. The last term within brackets on the right-hand side of Eq.~(\ref{eq:temperature}) results from the extended central mass distribution and is not present in the standard solution around a BH. For more details on the disk solution, including a plot showing the decreasing $T(r)$ behavior inside the DM core, we refer the reader to \citet{millauro2024}. 

Additionally, we assume blackbody emission in the local rest frame of the accretion disk at the temperature given in Eq.~(\ref{eq:temperature}). The rest-frame four-velocity is
\begin{equation}
    u^a_{\text{d}} \propto (\partial_t)^a + \Omega (\partial_\phi)^a\,,    
\end{equation}
where $\Omega$ is the angular velocity of Eq.~(\ref{eq:angular_speed}), and the four-velocity vector is subject to the normalization condition $g_{ab} u^{a}_{\mathrm{d}} u^{b}_{\mathrm{d}} = -1$.

Given the lack of an innermost stable circular orbit (ISCO) in the RAR spacetime, we consider two alternatives for the inner radius of the disk. The first alternative is the radius at which the accretion efficiency saturates to a maximum, $r_{\text{in}} = r_{\text{sat}}$ \citep{millauro2024}, with the relative error for the change in efficiency being equal to or lower than 1\%. This radius typically corresponds to $r_{\rm sat} \sim 0.1r_{\rm c}$, where $r_{\rm c}$ denotes the radius of the DM core. The second alternative assumes that the disk continues to the center (e.g., Figs.~\ref{fig:sch_bol} and \ref{fig:size_comparison}). We have quantitatively analyzed the minor differences among the images for these two scenarios in Sect.~\ref{sec:results} to demonstrate that our results are essentially insensitive to the choice of inner radius.

\section{Relativistic ray tracing}
\label{sec:methods}

Strong gravitational fields can significantly influence the observational features of compact objects through various effects, including light bending, gravitational redshift, and, if matter moves at considerable fractions of the speed of light, relativistic beaming. To accurately account for these effects, it is necessary to consider the covariant radiative transfer equation on a curved spacetime,
\begin{align}
    \frac{\mathop{d}}{\mathop{d\lambda}} \left( \frac{I_\nu}{\nu^3}\right) &= \frac{j_\nu}{\nu^2} - \nu \alpha_\nu \left( \frac{I_\nu}{\nu^3}\right) \,,
    \label{eq:complete_transfer}
\end{align}
where $\nu$ represents the radiation frequency, $I_\nu$ is the specific intensity of the radiation field, $j_\nu$ and $\alpha_\nu$ denote the emissivity and absorptivity coefficients of the medium, respectively, and $\lambda$ is an affine parameter along spacetime geodesics. In our model, the accretion disk is infinitesimally thin, with no emission or absorption processes occurring outside. Consequently, both plasma coefficients in the right-hand side of Eq.~(\ref{eq:complete_transfer}) are zero, and the radiative transfer is reduced to the connection between the emitting disk and the observation point along geodesics via the conservation of the Lorentz-invariant quantity $I_\nu/\nu^3$.  

For our imaging and spectral analysis, we utilize \texttt{skylight} \footnote{https://github.com/joaquinpelle/Skylight.jl}, an open-source Julia package for general-relativistic ray tracing and radiative transfer in arbitrary spacetimes \citep{pelle2022skylight}. \texttt{skylight} is particularly advantageous for its easy extensibility to custom spacetimes and radiative models, which we leverage to incorporate the RAR spacetime and accretion disk model into the code. Specifically, for the spacetime, we implement an interpolation scheme to compute the metric coefficients and Christoffel symbols using numerical data from the RAR solutions.

We utilize an updated version of \texttt{skylight}, which includes improvements over the description in \citet{pelle2022skylight}. Most importantly, it calculates observables not by relying on an image-plane approximation for distant observers but directly through the covariant energy-momentum tensor of the radiation field, 
\begin{equation}
    T^{ab} = \int k^a k^b (I_\nu/\nu^3) \nu d\nu \Omega\,, 
\end{equation}
where $k^a = \nu (1, \mathbf{\Omega})$, $\nu$ is the frequency, and $\mathbf{\Omega}$ is the spatial direction in an orthonormal frame. The flux calculation follows as $F = T^{ab} u_a n_b$, with $u^a$ being the four-velocity of the reference frame, and $n^a$ a vector normal to the flux-measurement surface, such that $n^a u_a = 0$ and $n^a n_a = 1$. This approach enables the precise calculation of fluxes from any position and frame, removing the need for asymptotic flatness of the spacetime. Our flux measurements correspond to the static frame along the radial direction. That is, we set $u^a \propto (\partial_t)^a$, and $n^a \propto (\partial_r)^a$, both with adequate normalization.

Finally, the multi-scale nature of the model poses a challenge for radiative transfer within the RAR halos. While the scale of the accretion disk is near that of the central object, the DM halo extends to several kiloparsecs. In principle, this requires integrating the equations across large distances while potentially considering spatial variations in much smaller scales, which can easily lead to numerical instabilities. However, if the influence of the DM halo on radiation is negligible beyond a certain distance, we can mitigate this issue by taking observation points relatively close to the source. Then, we can correct the flux and angle normalization for observations at larger distances according to an inverse-square law as in flat spacetime. In Appendix~\ref{app:distance}, we elaborate on the validity of taking our observation points at $r = 1 \, \text{pc}$ in all scenarios we consider in this work. 

\section{Model parameters}
\label{sec:parameters}

\begin{table*}
\centering
\sisetup{table-format=1.2e+1} 
\begin{tabular}{c | c c c c c}
\toprule
\quad $m \,$(keV) \quad & \quad $M_{\rm c}\,$($M_{\odot}$) \quad & \quad $r_g\,$(cm) \quad & \quad $r_{\text{c}}\,$($r_g$) \quad & \quad $r_{\text{sat}}\,$($r_g$) \quad & \quad $\theta_{\text{sat}}\,$(as)\footnote{The angular radius corresponding to $r_\text{sat}$ is calculated as $\theta_\text{sat} = r_\text{sat}/1\,\text{pc}$, where $1\,\text{pc}$ is the radius of the observation point.}  \\
\midrule
48  & \num{1e7} & \num{1.48e12} & \num{947} & \num{78.8} & 7.71 \\
155 & \num{1e7} & \num{1.48e12} & \num{36.7} & \num{3.09} & 0.30 \\
200 & \num{1e7} & \num{1.48e12} & \num{15.5} & \num{1.37} & 0.13 \\
200  & \num{3.5e6} & \num{5.17e11} & \num{81.6} & \num{6.87} & \num{0.23} \\
300 & \num{3.5e6} & \num{5.17e11} & \num{24.4} & \num{2.09} & 0.07 \\
378 & \num{3.5e6} & \num{5.17e11} & \num{7.84} & \num{0.71} & \num{0.02} \\
\bottomrule
\end{tabular}
\caption{Core mass $M_{\rm c}$, core gravitational radius $r_g$, core radius $r_{\text{c}}$, saturation radius $r_{\text{sat}}$, and corresponding angular radius $\theta_{\text{sat}}$ for the different RAR solutions considered in this work.}
\label{tab:rsat}
\end{table*}

We use two families of solutions within the RAR DM model. In Table~\ref{tab:rsat}, we show various parameters characterizing our chosen solutions. The first family corresponds to halos typical of active-like galaxies with a supermassive central object of mass $10^7 M_\odot$ and surrounded by a total halo mass of $10^{12} M_\odot$ (see Fig.~\ref{fig:rho}). In this case, we consider fermion masses of $m =48\,\text{keV}$, $155\,\text{keV}$, and $200\,\text{keV}$, in increasing order of core compactness. Thus, as explained in Sect. \ref{sec:model}, for a central object with $M_{\rm c}=10^7 M_\odot$, the $200\,$keV solution implies a fermion core close to its critical value of collapse into a SMBH. For the other solution, we adopt a halo profile with a smaller core mass close to $4\times 10^6 M_\odot$, surrounded by a Milky Way-like halo, as studied in \citet{2020A&A...641A..34B}. In this case, we first adopt a fermion mass of $200\,$keV, the same as in the case of the larger active galaxy. However, we also consider more compact cores with $m=300\,\text{keV}$ and $m=378\,\text{keV}$, the latter leading to the critical mass of core-collapse into a SMBH typical of a Milky Way-like galaxy.

The parameters of the accretion disk model include an accretion rate fixed at $10 \%$ of the Eddington accretion rate, with an efficiency of conversion to luminosity of $10 \%$ as well. As described in Sect.~\ref{sec:model}, we consider two alternatives for the inner radius of the disk: $r_\text{in} = 0$ and $r_\text{in} = r_\text{sat}$. We note that $r_\text{sat}$ depends on the DM distribution and, consequently, the fermion mass. We show the specific values of $r_\text{sat}$ in Table~\ref{tab:rsat} for each RAR solution considered here. We set the outer radius of the accretion disk to $10^3 r_\text{sat}$.

We take our observation points at $r=1\,\text{pc}$, for which we provide a detailed justification in Appendix~\ref{app:distance}. We take inclination angles relative to the rotation axis of the accretion disk of $\xi = 5^\circ$, $45^\circ$, $85^\circ$, that is, almost face-on, intermediate, and almost edge-on views, respectively. The cameras have angular apertures of $80 \theta_\text{sat}$, where $\theta_\text{sat} = r_\text{sat} / 1\,\text{pc}$ represents the angle subtended by the radius $r_\text{sat}$ from the observation point. For reference, we show the $\theta_\text{sat}$ values for each scenario in Table~\ref{tab:rsat}. We take camera resolutions of $1200 \times 1200$ pixels. 

For comparison, we also include simulations in a setup where the central object is a Schwarzschild BH with a mass of $M_{\rm BH} = 10^7 M_\odot$. In this model, we employ the standard Shakura-Sunyaev disk, extending down to the ISCO, with the same accretion rate as in the RAR scenario. The observation points are at the same location as in the RAR scenario. For the scale of the images, we adopt the value of $\theta_{\text{sat}}$ corresponding to the $200\,\text{keV}$ RAR solution with $M_{\rm c} = 10^7 M_\odot$.

\section{Results}
\label{sec:results}

\begin{figure*}
    \centering
    \includegraphics[width = \textwidth]{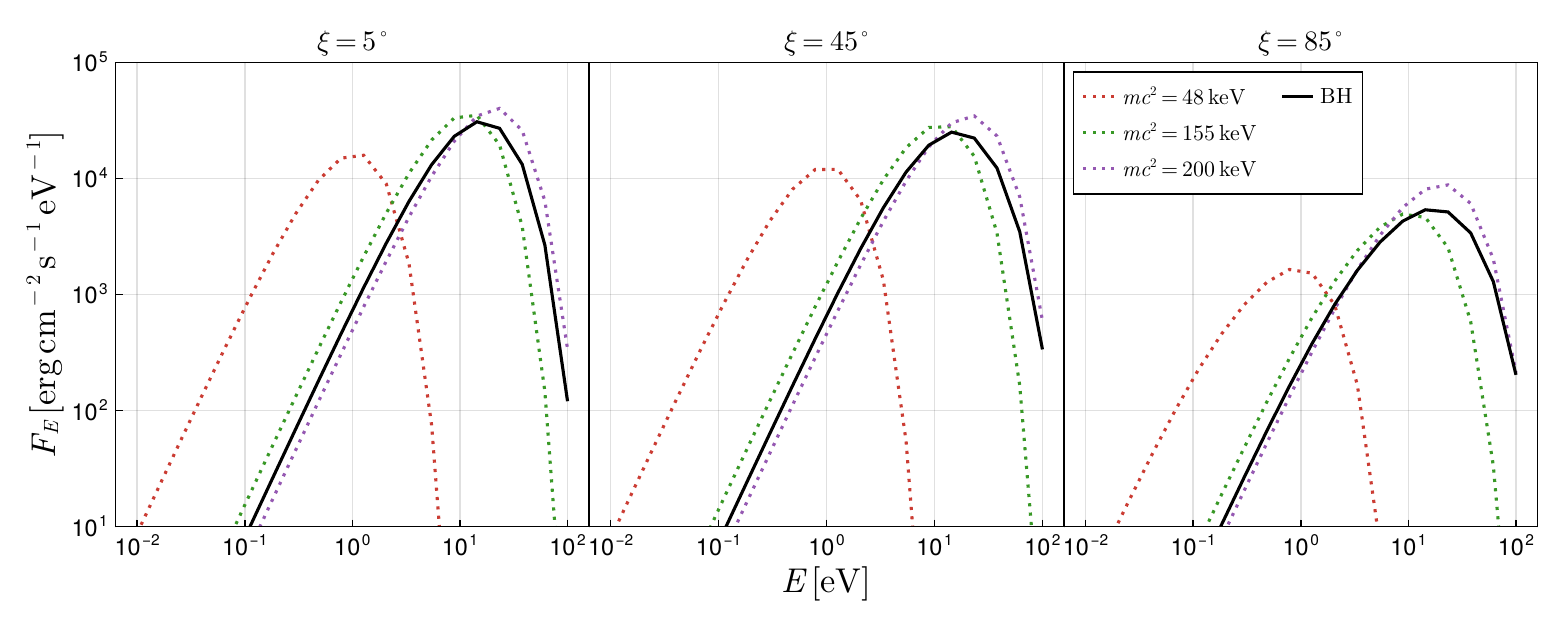}
    \caption{Specific fluxes for an accretion disk extending to the origin around a DM core of $M_{\rm c} = \num{1e7} M_\odot$. The columns correspond to observation inclinations relative to the rotation axis of the disk of $\xi = 5^\circ$, $45^\circ$, $85^\circ$. The flux is measured in the static frame along the radial direction.}   
    \label{fig:r0_spectrum}
\end{figure*}

In the following, we present the results, including images and spectra. In Fig.~\ref{fig:r0_spectrum}, we show the observed spectra at different inclination angles, comparing the RAR models with $r_\text{in}=0$ to the BH scenario. From the spectra, we can see that the more compact the DM core (i.e., the larger the fermion mass), the more luminous the solutions are overall, such that there exists a certain core compactness for which the emitted flux resembles that of a BH of the same mass as the DM core. Moreover, accretion onto the most compact RAR solution, which in this case corresponds to a fermion mass of 200 keV, can be more efficient and consequently result in higher fluxes than in the BH scenario (as obtained in \citet{millauro2024}). The fluxes in the RAR cases peak at higher frequencies for larger particle masses, corresponding to the higher temperatures of their disks \citetext{see \citealp{millauro2024} for a general discussion of this point}.       

\begin{figure*}
    \centering
    \includegraphics[width = \textwidth]{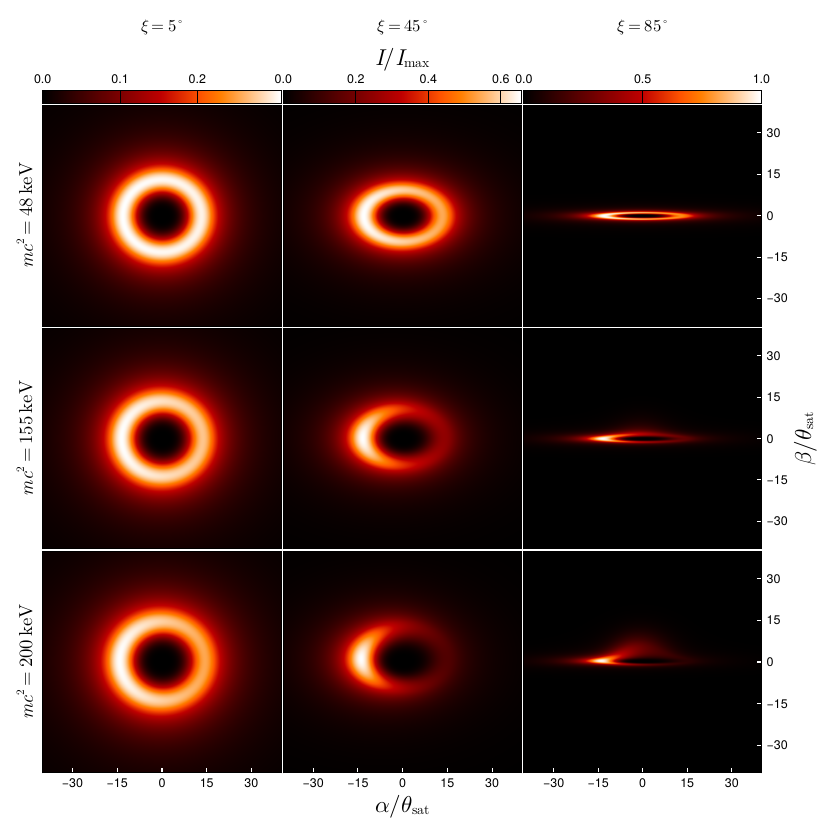}
    \caption{Bolometric intensity for an accretion disk extending to the origin around a DM core of $M_{\rm c} = \num{1e7} M_\odot$. The columns correspond to observation inclinations relative to the rotation axis of the disk of $\xi = 5^\circ$, $45^\circ$, $85^\circ$. The intensity is measured in the static frame. Here, $\alpha$ and $\beta$ denote the angular coordinates on the sky centered around the radial direction, and $\theta_\text{sat} = r_\text{sat}/r$, i.e. the angular radius corresponding to $r_\text{sat}$, which we tabulate in Table~\ref{tab:rsat}. The intensity has been scaled to $I_{\text{max}} = 9 \times 10^{15}   \, \text{erg} \, \text{cm}^{-2} \, \text{sr}^{-1} \, \text{s}^{-1}$.}
    \label{fig:r0_bol}
\end{figure*}

\begin{figure*}
    \centering
    \includegraphics[width = \textwidth]{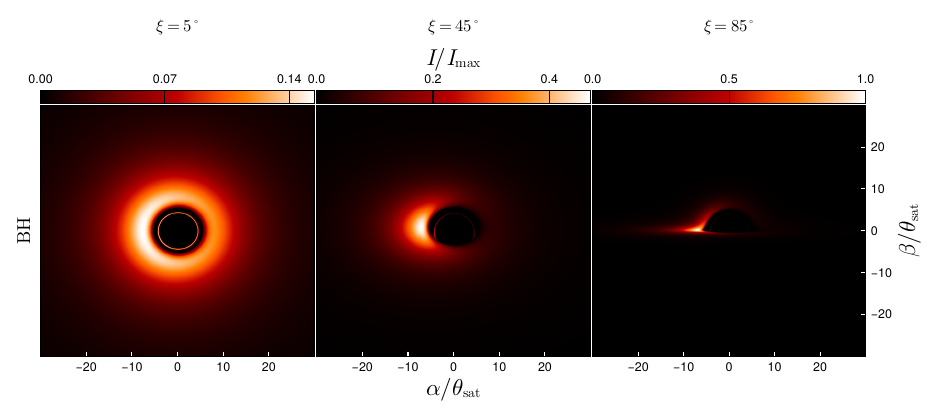}
    \caption{Bolometric intensity for an accretion disk around a BH of $M_{\rm BH} = \num{1e7} M_\odot$. The columns correspond to observation inclinations relative to the rotation axis of the disk of $\xi = 5^\circ$, $45^\circ$, $85^\circ$. The intensity is measured in the static frame. Here, $\alpha$ and $\beta$ denote the angular coordinates on the sky centered around the radial direction, and $\theta_\text{sat} = r_\text{sat}/r$, i.e., the angular radius corresponding to $r_\text{sat} = $ for the $200 \, \text{keV}$ configuration. The intensity has been scaled to $I_{\text{max}}= 1.8 \times 10^{16} \, \text{erg} \, \text{cm}^{-2} \, \text{sr}^{-1} \, \text{s}^{-1}$.}
    \label{fig:sch_bol}
\end{figure*}

In Fig.~\ref{fig:r0_bol}, we show bolometric images of the RAR solutions for the active galaxies (the first three solutions on Table~\ref{tab:rsat}). We show the same for the BH scenario in  Fig.~\ref{fig:sch_bol}. 

\begin{figure*}
    \centering
    \includegraphics[width = \textwidth]{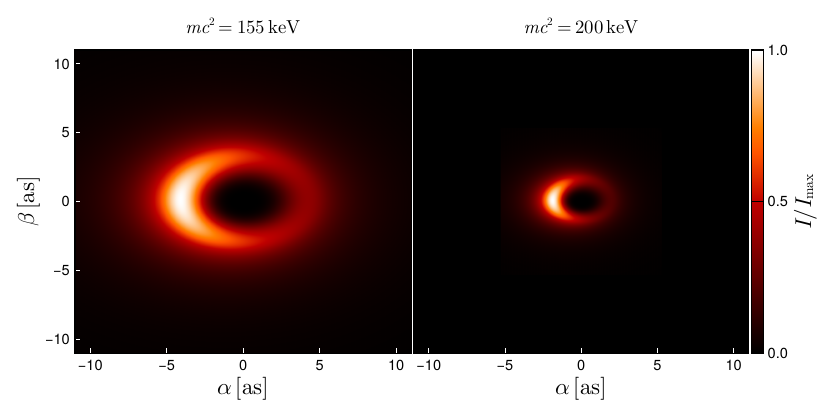}
    \caption{Bolometric intensity for an accretion disk extending to the origin around a DM core of $M_{\rm c} = \num{1e7} M_\odot$ with an observation inclination relative to the rotation axis of the disk of $\xi = 45^\circ$. The intensity is measured in the static frame. Here, $\alpha$ and $\beta$ denote the angular coordinates on the celestial sphere centered around the radial direction. The intensity has been scaled to $I_{\text{max}} = 6 \times 10^{15} \, \text{erg} \, \text{cm}^{-2} \, \text{sr}^{-1} \, \text{s}^{-1}$.}
    \label{fig:size_comparison}
\end{figure*}

In Fig.~\ref{fig:size_comparison}, we compare the different sizes of the solutions corresponding to $155$ and $200\,$keV for an inclination angle of $\xi=45^{\circ}$. Given its extension, the $48\,$keV solution, which is the least compact, is not included in the comparison.

\begin{figure*}
    \centering
    \includegraphics[width = \textwidth]{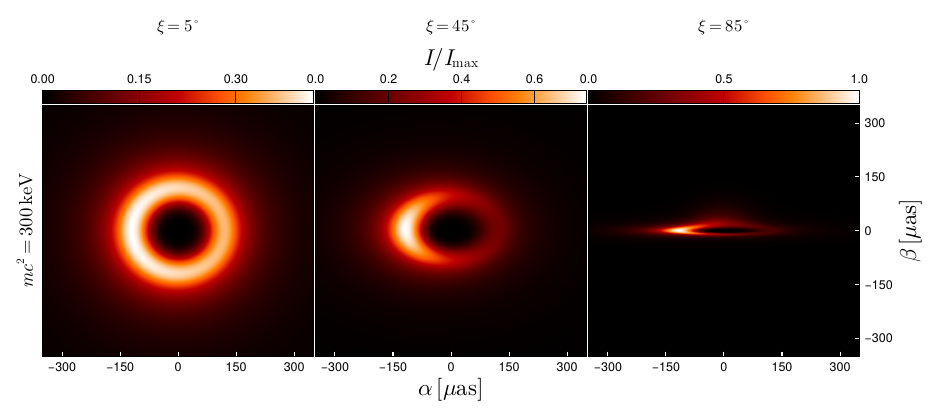}
    \caption{Bolometric intensity for an accretion disk with inner radius $r_{\text{in}}=r_{\text{sat}}$ around a DM core of $M_{\rm c} = \num{3.5e6} M_\odot$. The columns correspond to observation inclinations relative to the rotation axis of the disk of $\xi = 5^\circ$, $45^\circ$, $85^\circ$. The intensity is measured in the static frame. Here, $\alpha$ and $\beta$ denote the angular coordinates on the sky centered around the radial direction for a fictitious observer at $r=8.277\,\text{kpc}$. The intensity has been scaled to $I_{\text{max}}= 5.7 \times 10^{15} \, \text{erg} \, \text{cm}^{-2} \, \text{sr}^{-1} \, \text{s}^{-1}$.}
    \label{fig:mwlike_bol}
\end{figure*}

Figure~\ref{fig:mwlike_bol} shows the image for the Milky Way-like RAR solution for $m=300\,$keV with $r_\text{in}=r_\text{sat}$ (although we also considered other fermion masses of $200$ and $378\,$keV for this case, we do not include these images to avoid overload). Additionally, although the observation point for the numerical code is at $r=1\,\text{pc}$ as in every other case, we adopt a reference angular scale for the image corresponding to an observer at $r=8.277\,\text{kpc}$ (the approximate distance from Earth to Sgr A*), applying an effective decay of solid angles as in flat spacetime from the observation point of the code to the reference observer.

The images show differences in the peak brightness at different viewing angles due to enhanced relativistic beaming for more edge-on views. Furthermore, the edge-on BH image has a peak brightness almost twice as high as that in the RAR scenarios due to the higher angular velocities near the ISCO. As expected, the edge-on images exhibit moderate light bending for DM distributions with less compact cores. The effect becomes more pronounced for the more compact cores (as shown in Fig. \ref{fig:deflection}). Nonetheless, even in the latter cases, there are appreciable differences compared to the strong deflection seen in the BH scenario, which is characterized by severe image distortion of the disk portion behind the BH. Additionally, independent of the viewing angle, the BH scenario shows the presence of photon rings, whereas this is not the case for the RAR cores, not even in the critical core mass case, where the light-bending angle is always less than $\pi$ (see  Fig. \ref{fig:deflection}).
 
Asymmetric brightness patterns are present at intermediate and edge-on views due to relativistic beaming from the high velocities of matter along the line of sight near the core regions. This effect becomes more pronounced with increasing core compactness. Additionally, all the images show a central brightness depression surrounded by a ring-like structure primarily due to the temperature drop toward the center and, secondarily, to the redshift of photons from that region. 

Disk configurations with $r_\text{in}=0$ and $r_\text{in}=r_\text{sat}$ are almost visually indistinguishable. This similarity is because the temperatures peak at similar radii and drop abruptly toward the center in both cases. We confirmed the similarity quantitatively, with relative differences being around $0.1 \%$ on average at pixels where $r_0 (\alpha, \beta) \geq r_\text{sat}$, where $r_0(\alpha, \beta)$ is the radius of the fluid element observed at angular coordinates $(\alpha, \beta)$ on the image. Naturally, pixels where $r_0(\alpha, \beta) < r_\text{sat}$ have the highest contrast. However, between the $r_\text{in}=0$ and $r_\text{in}=r_\text{sat}$ configurations, the relative differences (under the $L_2$-norm) of the observed overall spectra are below $0.1 \%$ in all scenarios we consider.

\section{Discussion}\label{sec:discussion}

We have calculated the synthetic images produced by accretion disks onto fermionic DM cores at different viewing angles. We have considered a geometrically thin and optically thick disk model with a composite black body spectrum given by a radius-dependent temperature profile \citep{millauro2024}. We applied our model to active-like galaxies with core masses of $10^7 M_{\odot}$, and using three different DM particle masses: $48$, $155$, and $200\,$keV, which result in different core compactness and consequently different image sizes. Additionally, we studied a hypothetical Milky Way-like galaxy with a lighter DM core of approximately $4\times 10^6 M_\odot$ for different fermion masses of $200, 300$, and $378\,$keV, implying different core compactness, where, in the latter case, the core is close to its critical mass for gravitational collapse into a BH \citep{arguelles2018}. We discuss the main results in the following sections.

\subsection{Central brightness depression}

A notable difference between the DM core and BHs is the absence of an ISCO in the RAR solutions. Spherical fermionic cores admit stable circular orbits at all radii \citep{crespi2022estudio}. Therefore, in our model, matter can, at least in principle, plunge toward the central region while emitting radiation from the friction between differentially rotating layers\footnote{As a side note, we mention that in the BH scenario, theoretical and observational analyses have indicated that matter can significantly radiate when plunging between the ISCO and the event horizon \citep{schnittman2016disk, mummery2024continuum, mummery2024plunging}.}. The central brightness depression and surrounding ring-like feature in the images for the RAR solutions arise from the temperature drop toward the center due to the decay of viscous stresses (see also Eq.~\ref{eq:temperature}), and, secondarily, from the redshift of photons coming from the central region within the core. Thus, we do not expect the inner edge of the central brightness depression in the RAR case to be as sharp as in the BH case with a thin accretion disk truncated at the ISCO.

\subsection{Photon rings}    

\begin{figure}
    \centering
    \includegraphics[width = \columnwidth]{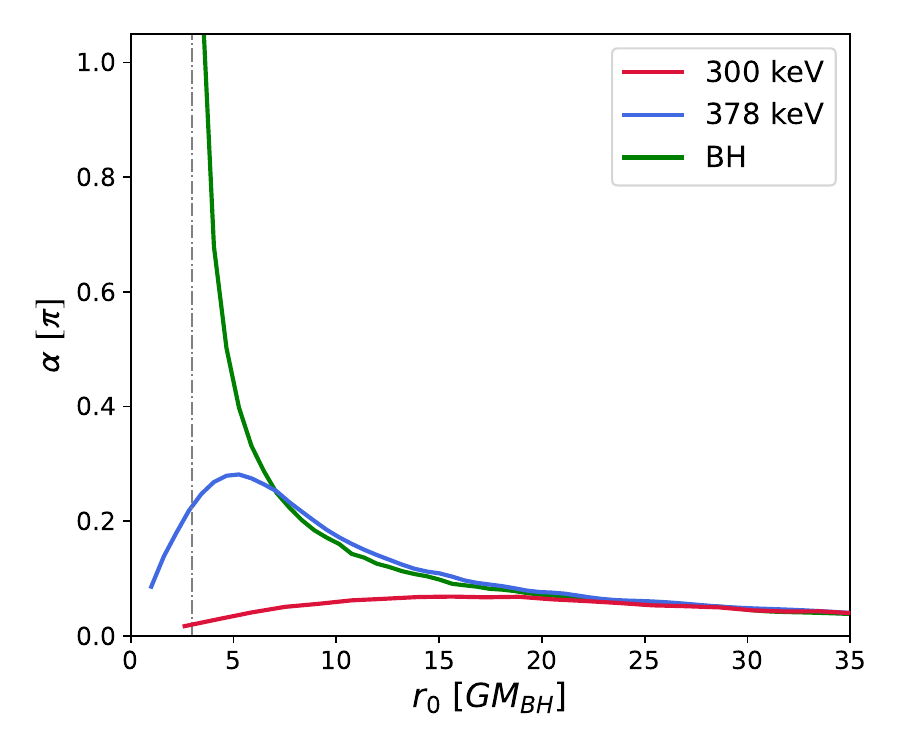}
    \caption{Deflection angle for massless particles in the RAR (blue and red curves), and  Schwarzschild (green line) spacetimes, as a function of closest approach to the center. $M_{{\rm BH}}=4\times 10^6 M_\odot$.}   
    \label{fig:deflection}
\end{figure}

As with any static spherically symmetric perfect fluid solution within GR under a fermionic equation of state, the RAR solutions do not have photon rings. The most compact (i.e., critical) solution for self-gravitating neutral massive (spin $1/2$) fermions corresponds to full degeneracy, with $R_{\rm OV}/M_{\rm OV}=8.8$ \citep[e.g.,][]{2020EPJB...93..208A}, where $R_{\rm OV}$ and $M_{\rm OV}$ are the Oppenheimer-Volkoff radius and mass, respectively. This lack of photon rings contrasts with the case of BHs, for which photon rings are a universal prediction \citep{gralla2019black, gralla2020lensing, johnson2020universal}. The maximum deflection for massless particles in the RAR spacetimes we consider is approximately $3\pi/10$ (\citealt{2016PhRvD..94l3004G}; see Fig. \ref{fig:deflection} for the specific examples treated in this article), corresponding to the most compact degenerate fermion core with $r_c/M_{\rm c}\approx 9$. This bound implies that photons do not produce secondary (or higher-order) images due to the lack of null geodesics with multiple plane crossings. 

The observation of photon rings around the supermassive objects at the center of our Galaxy and M87 may soon become possible \citep{johnson2020universal,younsi2023black,kocherlakota2024hotspots}. Despite the enhanced capabilities of the next-generation Event Horizon Telescope \citep[ngEHT,][]{johnson2023key,ayzenberg2023fundamental}, measuring photon rings solely with Earth-based interferometry will continue to be challenging even in the ultra-high resolution regime due to the dynamic nature, changing opacity, and rapid variability of these environments \citep{tiede2022measuring}. However, future arrays that integrate space-based instruments extending the VLBI baseline are expected to enable these measurements \citep{gurvits2022science, johnson2024bhex}. If realized, these would discriminate between the fermionic DM model and the BH paradigm.

\subsection{Imaging a $4\times 10^6 M_\odot$ fermion core}

In addition, we also calculated the image for a disk in a galaxy with the same core mass as that of Sgr A*. For this, we used core--halo RAR solutions with particle masses of $200, 300$, and $378\,$keV, such that in the latter case, the highly degenerate fermion core reaches the Oppenheimer-Volkoff critical mass for gravitational collapse into a BH of about $4\times10^6 M_\odot$, with the outer halo explaining the Milky Way rotation curve \citep{arguelles2018}. 

We emphasize, however, that our accretion model is not suitable for Sgr A*, as the observed spectral characteristics are consistent with synchrotron radiation from a thick, hot disk with a lower accretion rate rather than thermal blackbody radiation from a thin disk, as shown in Fig.~\ref{fig:mwlike_bol}. That said, we obtain global features that are at least promising for future work. Specifically, the diameter of the central brightness depression for the $m=300\,$keV example shown in Fig. \ref{fig:mwlike_bol}, as seen from $r=8.277\,\text{kpc}$, is approximately $100\,\mu\mathrm{as}$ for intermediate viewing angles. In contrast, for the $m=200\,$keV case, the diameter is close to three times larger, with a value of approximately $300\,\mu\mathrm{as}$. Diameters of the central dark region around that of the EHT image of Sgr A* \citetext{$48.7 \pm 7\,\mu\mathrm{as}$, \citealp{akiyama2022-Ifirst}} can be obtained for more compact DM core solutions than the $m=300\,$keV case, though not as compact as the critical fermion core solution. Indeed, such a critical solution with a fermion mass of $378\,$keV and a core mass of $3.5 \times 10^6 M_\odot$ has a core radius of $7.85 r_g$, and a central dark region diameter of approximately $35\,\mu\mathrm{as}$, about three times smaller than that of the $300\,$keV solution. In future work, we aim to develop a more suitable accretion disk model around RAR fermion cores for low accretion rates, such as those estimated for the Sgr A*. This development will allow us to compare the resulting images more accurately with those obtained by the EHT.

\subsection{Absence of an event horizon}

A defining characteristic of BH solutions is the presence of an event horizon. However, only apparent horizons are physically detectable by quasi-local measurements \citetext{see \citealp{2014PhRvD..90l7502V} and references therein}. Early in the development of the VLBI technology associated with the EHT, it was proposed that detecting a shadow would provide evidence for the existence of an event horizon \citep{2000ApJ...528L..13F}. Since then, many works have investigated shadow features for different BH spacetimes and horizonless compact object alternatives within and beyond GR \citetext{see \citealp{2022PhR...947....1P} for a review}. 

A broad family of BH mimickers is the so-called ultra-compact horizonless objects \citep{2018PhRvD..98l4009C}, characterized by a physical surface lying just above the would-be event horizon and within the photon surface of the BH candidate. Stimulated by observational results that claimed evidence for the existence of event horizons at galaxy centers \citep{2004ApJ...615..402M,2008NewAR..51..733N}, many works have put constraints on such BH alternatives from the non-observation of electromagnetic radiation from the putative surface of such objects \citetext{see \citealp{2022ApJ...930L..17E,2022JCAP...08..055C} for recent results on Sgr A*}. However, \citet{2014PhRvD..90l7502V} argued that the observation of the accretion flows analyzed in \citet{2008NewAR..51..733N} only suggests the presence of some sort of horizon (i.e., a one-way membrane of infalling matter), but cannot be interpreted as a claim for the presence of an event horizon. 

Such an observable condition of `no emitting hard surface' is fulfilled by specific BH alternatives, such as compact objects made of bosons or fermions. The first case corresponds to the well-studied boson stars \citetext{e.g., \citealp{vincent2016} in the context of the present discussion}, while the second case, and the subject of this paper, corresponds to highly degenerate DM fermion cores surrounded by a DM halo made of the same particles \citep{arguelles2018,2019PDU....24..278A}.   

\section{Conclusions}
\label{sec:conclusions}

We have presented the images cast by an accretion disk around nonsingular, horizonless, fermionic DM cores at the center of active-like galaxies. The neutral (spin $1/2$) massive fermions constituting the core are in a dense and degenerate state, followed by a sharp decrease in density ending in a diluted atmosphere that explains the DM halos in such galaxies. These core--halo DM distributions are known as the RAR profiles and are solutions of the Einstein field equations in spherical symmetry sourced by a perfect fluid ansatz for the finite temperature fermions \citep{2015MNRAS.451..622R,arguelles2018,2019PDU....24..278A,2021MNRAS.502.4227A}. For particle masses of the order of $50-350\,$keV, such RAR fermion cores have recently been proven to be an alternative to the BH paradigm in Sgr A* since they explain the orbits of the S-star cluster \citep{2020A&A...641A..34B,2021MNRAS.505L..64B,2022MNRAS.511L..35A}, while acting as SMBH alternatives in larger galaxies (or eventually collapsing into one) \citep{2019PDU....24..278A,2024ApJ...961L..10A}. We have extended those former results by studying the relativistic images produced by the lensed photons originating from thin disks around the fermionic cores, complementing analogous results already shown for boson stars in \citet{2006PhRvD..73b1501G,vincent2016}, though in our case applied to typical active galaxies.

We have obtained image features similar to those expected in the BH case of the same mass as the DM core, with the central brightness depression and surrounding ring-like structure being the most relevant. However, we also find differences, such as the absence of photon rings in the fermionic case. For the case of a Milky Way-like galaxy with a fermion core of about $4\times 10^6 M_\odot$, we found relativistic images with central brightness depressions consistent with those observed by the EHT. However, we emphasize that our accretion model is unsuitable for Sgr A*. Further work will be devoted to developing an accretion model that applies to Sgr A*, given the improved interferometric capabilities and resolutions that will be possible in the near future \citep{johnson2023key,ayzenberg2023fundamental,gurvits2022science,johnson2024bhex}, especially including space-based instruments.

\section*{Acknowledgements}
J.P. acknowledges financial support from a CONICET fellowship. C.R.A. acknowledges support from the Argentine agency CONICET (PIP2876), the ANPCyT (grant PICT-2020-02990), and ICRANet. F.L.V. acknowledges support from the Argentine agency CONICET (PIP 2021-0554), and the ANPCyT (PICT 2022).
M.F.M. acknowledges support from CONICET (PIP2169) and the Universidad Nacional de La Plata (PID G178). This work used computational resources from CCAD—Universidad Nacional de Córdoba (https://ccad.unc.edu.ar/), which is part of SNCAD—MinCyT, República Argentina. We would like to acknowledge the collaboration opportunities that arose from the Friends of Friends Meeting in Argentina (\url{https://fof.oac.uncor.edu/2022/}), which originated this work. We produced the visualizations of \texttt{skylight} data using \textit{Makie.jl} \citep{danisch2021makie}, a package for high-performance data visualization in Julia.

\section*{Data Availability}

The data underlying this article will be shared on reasonable request to the corresponding author.



\bibliographystyle{mnras}
\bibliography{main} 

\begin{thebibliography}{}
\makeatletter
\relax
\def\mn@urlcharsother{\let\do\@makeother \do\$\do\&\do\#\do\^\do\_\do\%\do\~}
\def\mn@doi{\begingroup\mn@urlcharsother \@ifnextchar [ {\mn@doi@} {\mn@doi@[]}}
\def\mn@doi@[#1]#2{\def\@tempa{#1}\ifx\@tempa\@empty \href {http://dx.doi.org/#2} {doi:#2}\else \href {http://dx.doi.org/#2} {#1}\fi \endgroup}
\def\mn@eprint#1#2{\mn@eprint@#1:#2::\@nil}
\def\mn@eprint@arXiv#1{\href {http://arxiv.org/abs/#1} {{\tt arXiv:#1}}}
\def\mn@eprint@dblp#1{\href {http://dblp.uni-trier.de/rec/bibtex/#1.xml} {dblp:#1}}
\def\mn@eprint@#1:#2:#3:#4\@nil{\def\@tempa {#1}\def\@tempb {#2}\def\@tempc {#3}\ifx \@tempc \@empty \let \@tempc \@tempb \let \@tempb \@tempa \fi \ifx \@tempb \@empty \def\@tempb {arXiv}\fi \@ifundefined {mn@eprint@\@tempb}{\@tempb:\@tempc}{\expandafter \expandafter \csname mn@eprint@\@tempb\endcsname \expandafter{\@tempc}}}

\bibitem[\protect\citeauthoryear{Akiyama et~al.,}{Akiyama et~al.}{2022a}]{akiyama2022-Ifirst}
Akiyama K.,  et~al., 2022a, The Astrophysical Journal Letters, 930, L12

\bibitem[\protect\citeauthoryear{Akiyama et~al.,}{Akiyama et~al.}{2022b}]{akiyama2022first}
Akiyama K.,  et~al., 2022b, The Astrophysical Journal Letters, 930, L13

\bibitem[\protect\citeauthoryear{Akiyama et~al.,}{Akiyama et~al.}{2022c}]{akiyama2022-IIIfirst}
Akiyama K.,  et~al., 2022c, The Astrophysical Journal Letters, 930, L14

\bibitem[\protect\citeauthoryear{Akiyama et~al.,}{Akiyama et~al.}{2022d}]{akiyama2022-IVfirst}
Akiyama K.,  et~al., 2022d, The Astrophysical Journal Letters, 930, L15

\bibitem[\protect\citeauthoryear{{Alberti} \& {Chavanis}}{{Alberti} \& {Chavanis}}{2020}]{2020EPJB...93..208A}
{Alberti} G.,  {Chavanis} P.-H.,  2020, \mn@doi [European Physical Journal B] {10.1140/epjb/e2020-100557-6}, \href {https://ui.adsabs.harvard.edu/abs/2020EPJB...93..208A} {93, 208}

\bibitem[\protect\citeauthoryear{{Arg{\"u}elles}, {Mavromatos}, {Rueda}  \& {Ruffini}}{{Arg{\"u}elles} et~al.}{2016}]{2016JCAP...04..038A}
{Arg{\"u}elles} C.~R.,  {Mavromatos} N.~E.,  {Rueda} J.~A.,   {Ruffini} R.,  2016, \mn@doi [J. Cosmol. Astropart. Phys.] {10.1088/1475-7516/2016/04/038}, \href {http://adsabs.harvard.edu/abs/2016JCAP...04..038A} {4, 038}

\bibitem[\protect\citeauthoryear{{Arg{\"u}elles}, {Krut}, {Rueda}  \& {Ruffini}}{{Arg{\"u}elles} et~al.}{2018}]{arguelles2018}
{Arg{\"u}elles} C.~R.,  {Krut} A.,  {Rueda} J.~A.,   {Ruffini} R.,  2018, \mn@doi [Physics of the Dark Universe] {10.1016/j.dark.2018.07.002}, \href {https://ui.adsabs.harvard.edu/abs/2018PDU....21...82A} {21, 82}

\bibitem[\protect\citeauthoryear{{Arg{\"u}elles}, {Krut}, {Rueda}  \& {Ruffini}}{{Arg{\"u}elles} et~al.}{2019a}]{2019PDU....24..278A}
{Arg{\"u}elles} C.~R.,  {Krut} A.,  {Rueda} J.~A.,   {Ruffini} R.,  2019a, \mn@doi [Physics of the Dark Universe] {https://doi.org/10.1016/j.dark.2019.100278}, \href {https://ui.adsabs.harvard.edu/abs/2019PDU....24..278A} {24, 100278}

\bibitem[\protect\citeauthoryear{{Arg{\"u}elles}, {Krut}, {Rueda}  \& {Ruffini}}{{Arg{\"u}elles} et~al.}{2019b}]{2019IJMPD..2843003A}
{Arg{\"u}elles} C.~R.,  {Krut} A.,  {Rueda} J.~A.,   {Ruffini} R.,  2019b, \mn@doi [International Journal of Modern Physics D] {10.1142/S021827181943003X}, \href {https://ui.adsabs.harvard.edu/abs/2019IJMPD..2843003A} {28, 1943003}

\bibitem[\protect\citeauthoryear{{Arg{\"u}elles}, {D{\'\i}az}, {Krut}  \& {Yunis}}{{Arg{\"u}elles} et~al.}{2021}]{2021MNRAS.502.4227A}
{Arg{\"u}elles} C.~R.,  {D{\'\i}az} M.~I.,  {Krut} A.,   {Yunis} R.,  2021, \mn@doi [Monthly Notices of the Royal Astronomical Society] {10.1093/mnras/staa3986}, \href {https://ui.adsabs.harvard.edu/abs/2021MNRAS.502.4227A} {502, 4227}

\bibitem[\protect\citeauthoryear{{Arg{\"u}elles}, {Becerra-Vergara}, {Krut}, {Yunis}, {Rueda}  \& {Ruffini}}{{Arg{\"u}elles} et~al.}{2022a}]{2022IJMPD..3130002A}
{Arg{\"u}elles} C.~R.,  {Becerra-Vergara} E.~A.,  {Krut} A.,  {Yunis} R.,  {Rueda} J.~A.,   {Ruffini} R.,  2022a, \mn@doi [International Journal of Modern Physics D] {10.1142/S0218271822300026}, \href {https://ui.adsabs.harvard.edu/abs/2022IJMPD..3130002A} {31, 2230002}

\bibitem[\protect\citeauthoryear{{Arg{\"u}elles}, {Mestre}, {Becerra-Vergara}, {Crespi}, {Krut}, {Rueda}  \& {Ruffini}}{{Arg{\"u}elles} et~al.}{2022b}]{2022MNRAS.511L..35A}
{Arg{\"u}elles} C.~R.,  {Mestre} M.~F.,  {Becerra-Vergara} E.~A.,  {Crespi} V.,  {Krut} A.,  {Rueda} J.~A.,   {Ruffini} R.,  2022b, \mn@doi [Monthly Notices of the Royal Astronomical Society] {10.1093/mnrasl/slab126}, \href {https://ui.adsabs.harvard.edu/abs/2022MNRAS.511L..35A} {511, L35}

\bibitem[\protect\citeauthoryear{{Arg{\"u}elles}, {Becerra-Vergara}, {Rueda}  \& {Ruffini}}{{Arg{\"u}elles} et~al.}{2023a}]{2023Univ....9..197A}
{Arg{\"u}elles} C.~R.,  {Becerra-Vergara} E.~A.,  {Rueda} J.~A.,   {Ruffini} R.,  2023a, \mn@doi [Universe] {10.3390/universe9040197}, \href {https://ui.adsabs.harvard.edu/abs/2023Univ....9..197A} {9, 197}

\bibitem[\protect\citeauthoryear{{Arg{\"u}elles}, {Boshkayev}, {Krut}, {Nurbakhyt}, {Rueda}, {Ruffini}, {Uribe-Su{\'a}rez}  \& {Yunis}}{{Arg{\"u}elles} et~al.}{2023b}]{2023MNRAS.523.2209A}
{Arg{\"u}elles} C.~R.,  {Boshkayev} K.,  {Krut} A.,  {Nurbakhyt} G.,  {Rueda} J.~A.,  {Ruffini} R.,  {Uribe-Su{\'a}rez} J.~D.,   {Yunis} R.,  2023b, \mn@doi [Monthly Notices of the Royal Astronomical Society] {10.1093/mnras/stad1380}, \href {https://ui.adsabs.harvard.edu/abs/2023MNRAS.523.2209A} {523, 2209}

\bibitem[\protect\citeauthoryear{{Arg{\"u}elles}, {Rueda}  \& {Ruffini}}{{Arg{\"u}elles} et~al.}{2024}]{2024ApJ...961L..10A}
{Arg{\"u}elles} C.~R.,  {Rueda} J.~A.,   {Ruffini} R.,  2024, \mn@doi [\apj Letters] {10.3847/2041-8213/ad1490}, \href {https://ui.adsabs.harvard.edu/abs/2024ApJ...961L..10A} {961, L10}

\bibitem[\protect\citeauthoryear{Ayzenberg et~al.,}{Ayzenberg et~al.}{2023}]{ayzenberg2023fundamental}
Ayzenberg D.,  et~al., 2023, arXiv preprint arXiv:2312.02130

\bibitem[\protect\citeauthoryear{{Becerra-Vergara}, {Arg{\"u}elles}, {Krut}, {Rueda}  \& {Ruffini}}{{Becerra-Vergara} et~al.}{2020}]{2020A&A...641A..34B}
{Becerra-Vergara} E.~A.,  {Arg{\"u}elles} C.~R.,  {Krut} A.,  {Rueda} J.~A.,   {Ruffini} R.,  2020, \mn@doi [Astronomy \& Astrophysics] {10.1051/0004-6361/201935990}, \href {https://ui.adsabs.harvard.edu/abs/2020A&A...641A..34B} {641, A34}

\bibitem[\protect\citeauthoryear{{Becerra-Vergara}, {Arg{\"u}elles}, {Krut}, {Rueda}  \& {Ruffini}}{{Becerra-Vergara} et~al.}{2021}]{2021MNRAS.505L..64B}
{Becerra-Vergara} E.~A.,  {Arg{\"u}elles} C.~R.,  {Krut} A.,  {Rueda} J.~A.,   {Ruffini} R.,  2021, \mn@doi [Monthly Notices of the Royal Astronomical Society] {10.1093/mnrasl/slab051}, \href {https://ui.adsabs.harvard.edu/abs/2021MNRAS.505L..64B} {505, L64}

\bibitem[\protect\citeauthoryear{Boero \& Moreschi}{Boero \& Moreschi}{2021}]{boero2021strong}
Boero E.~F.,  Moreschi O.~M.,  2021, \mn@doi [Monthly Notices of the Royal Astronomical Society] {10.1093/mnras/stab2336}, 507, 5974

\bibitem[\protect\citeauthoryear{{Carballo-Rubio}, {Di Filippo}, {Liberati}  \& {Visser}}{{Carballo-Rubio} et~al.}{2018}]{2018PhRvD..98l4009C}
{Carballo-Rubio} R.,  {Di Filippo} F.,  {Liberati} S.,   {Visser} M.,  2018, \mn@doi [\prd] {10.1103/PhysRevD.98.124009}, \href {https://ui.adsabs.harvard.edu/abs/2018PhRvD..98l4009C} {98, 124009}

\bibitem[\protect\citeauthoryear{{Carballo-Rubio}, {Di Filippo}, {Liberati}  \& {Visser}}{{Carballo-Rubio} et~al.}{2022}]{2022JCAP...08..055C}
{Carballo-Rubio} R.,  {Di Filippo} F.,  {Liberati} S.,   {Visser} M.,  2022, \mn@doi [J. Cosmol. Astropart. Phys.] {10.1088/1475-7516/2022/08/055}, \href {https://ui.adsabs.harvard.edu/abs/2022JCAP...08..055C} {2022, 055}

\bibitem[\protect\citeauthoryear{{Cardoso} \& {Pani}}{{Cardoso} \& {Pani}}{2019}]{cardoso2019}
{Cardoso} V.,  {Pani} P.,  2019, \mn@doi [Living Reviews in Relativity] {10.1007/s41114-019-0020-4}, \href {https://ui.adsabs.harvard.edu/abs/2019LRR....22....4C} {22, 4}

\bibitem[\protect\citeauthoryear{Carroll}{Carroll}{2019}]{carroll2019spacetime}
Carroll S.~M.,  2019, Spacetime and geometry.
Cambridge University Press

\bibitem[\protect\citeauthoryear{{Cattoen}, {Faber}  \& {Visser}}{{Cattoen} et~al.}{2005}]{cattoen2005}
{Cattoen} C.,  {Faber} T.,   {Visser} M.,  2005, \mn@doi [Classical and Quantum Gravity] {10.1088/0264-9381/22/20/002}, \href {https://ui.adsabs.harvard.edu/abs/2005CQGra..22.4189C} {22, 4189}

\bibitem[\protect\citeauthoryear{Chavanis}{Chavanis}{1998}]{chavanis1998}
Chavanis P.-H.,  1998, Monthly Notices of the Royal Astronomical Society, 300, 981

\bibitem[\protect\citeauthoryear{{Chavanis}}{{Chavanis}}{2022}]{2022PhRvD.106d3538C}
{Chavanis} P.-H.,  2022, \mn@doi [\prd] {10.1103/PhysRevD.106.043538}, \href {https://ui.adsabs.harvard.edu/abs/2022PhRvD.106d3538C} {106, 043538}

\bibitem[\protect\citeauthoryear{{Chavanis} \& {Alberti}}{{Chavanis} \& {Alberti}}{2020}]{CHAVANIS2020135155}
{Chavanis} P.-H.,  {Alberti} G.,  2020, \mn@doi [Physics Letters B] {https://doi.org/10.1016/j.physletb.2019.135155}, 801, 135155

\bibitem[\protect\citeauthoryear{Crespi}{Crespi}{2022}]{crespi2022estudio}
Crespi V.,  2022, M.sc. thesis, UNLP. Available at \href{http://sedici.unlp.edu.ar/handle/10915/133910}{SIDECI}

\bibitem[\protect\citeauthoryear{Danisch \& Krumbiegel}{Danisch \& Krumbiegel}{2021}]{danisch2021makie}
Danisch S.,  Krumbiegel J.,  2021, Journal of Open Source Software, 6, 3349

\bibitem[\protect\citeauthoryear{{Do} et~al.,}{{Do} et~al.}{2019}]{2019Sci...365..664D}
{Do} T.,  et~al., 2019, \mn@doi [Science] {10.1126/science.aav8137}, \href {https://ui.adsabs.harvard.edu/abs/2019Sci...365..664D} {365, 664}

\bibitem[\protect\citeauthoryear{{Event Horizon Telescope Collaboration} et~al.,}{{Event Horizon Telescope Collaboration} et~al.}{2019}]{2019ApJ...875L...1E}
{Event Horizon Telescope Collaboration} et~al., 2019, \mn@doi [\apj Letters] {10.3847/2041-8213/ab0ec7}, \href {https://ui.adsabs.harvard.edu/abs/2019ApJ...875L...1E} {875, L1}

\bibitem[\protect\citeauthoryear{{Event Horizon Telescope Collaboration} et~al.,}{{Event Horizon Telescope Collaboration} et~al.}{2022a}]{2022ApJ...930L..12E}
{Event Horizon Telescope Collaboration} et~al., 2022a, \mn@doi [\apj Letters] {10.3847/2041-8213/ac6674}, \href {https://ui.adsabs.harvard.edu/abs/2022ApJ...930L..12E} {930, L12}

\bibitem[\protect\citeauthoryear{{Event Horizon Telescope Collaboration} et~al.,}{{Event Horizon Telescope Collaboration} et~al.}{2022b}]{2022ApJ...930L..17E}
{Event Horizon Telescope Collaboration} et~al., 2022b, \mn@doi [\apj Letters] {10.3847/2041-8213/ac6756}, \href {https://ui.adsabs.harvard.edu/abs/2022ApJ...930L..17E} {930, L17}

\bibitem[\protect\citeauthoryear{{Falcke}, {Melia}  \& {Agol}}{{Falcke} et~al.}{2000}]{2000ApJ...528L..13F}
{Falcke} H.,  {Melia} F.,   {Agol} E.,  2000, \mn@doi [\apj Letters] {10.1086/312423}, \href {https://ui.adsabs.harvard.edu/abs/2000ApJ...528L..13F} {528, L13}

\bibitem[\protect\citeauthoryear{{Ferrarese}}{{Ferrarese}}{2002}]{ferrarese2002}
{Ferrarese} L.,  2002, \mn@doi [\apj] {10.1086/342308}, \href {https://ui.adsabs.harvard.edu/abs/2002ApJ...578...90F} {578, 90}

\bibitem[\protect\citeauthoryear{{GRAVITY Collaboration} et~al.,}{{GRAVITY Collaboration} et~al.}{2018}]{2018A&A...615L..15G}
{GRAVITY Collaboration} et~al., 2018, \mn@doi [Astronomy $\&$ Astrophysics] {10.1051/0004-6361/201833718}, \href {https://ui.adsabs.harvard.edu/abs/2018A&A...615L..15G} {615, L15}

\bibitem[\protect\citeauthoryear{{GRAVITY Collaboration} et~al.,}{{GRAVITY Collaboration} et~al.}{2020}]{2020A&A...636L...5G}
{GRAVITY Collaboration} et~al., 2020, \mn@doi [Astronomy $\&$ Astrophysics] {10.1051/0004-6361/202037813}, \href {https://ui.adsabs.harvard.edu/abs/2020A&A...636L...5G} {636, L5}

\bibitem[\protect\citeauthoryear{{Genzel}, {Eisenhauer}  \& {Gillessen}}{{Genzel} et~al.}{2010}]{2010RvMP...82.3121G}
{Genzel} R.,  {Eisenhauer} F.,   {Gillessen} S.,  2010, \mn@doi [Rev. Mod. Phys.] {10.1103/RevModPhys.82.3121}, \href {http://adsabs.harvard.edu/abs/2010RvMP...82.3121G} {82, 3121}

\bibitem[\protect\citeauthoryear{{Ghez}, {Salim}, {Hornstein}, {Tanner}, {Lu}, {Morris}, {Becklin}  \& {Duch{\^e}ne}}{{Ghez} et~al.}{2005}]{2005ApJ...620..744G}
{Ghez} A.~M.,  {Salim} S.,  {Hornstein} S.~D.,  {Tanner} A.,  {Lu} J.~R.,  {Morris} M.,  {Becklin} E.~E.,   {Duch{\^e}ne} G.,  2005, \mn@doi [\apj] {10.1086/427175}, \href {https://ui.adsabs.harvard.edu/abs/2005ApJ...620..744G} {620, 744}

\bibitem[\protect\citeauthoryear{{Ghez} et~al.,}{{Ghez} et~al.}{2008}]{2008ApJ...689.1044G}
{Ghez} A.~M.,  et~al., 2008, \mn@doi [\apj] {10.1086/592738}, \href {http://adsabs.harvard.edu/abs/2008ApJ...689.1044G} {689, 1044}

\bibitem[\protect\citeauthoryear{{G{\'o}mez}, {Arg{\"u}elles}, {Perlick}, {Rueda}  \& {Ruffini}}{{G{\'o}mez} et~al.}{2016}]{2016PhRvD..94l3004G}
{G{\'o}mez} L.~G.,  {Arg{\"u}elles} C.~R.,  {Perlick} V.,  {Rueda} J.~A.,   {Ruffini} R.,  2016, \mn@doi [\prd] {10.1103/PhysRevD.94.123004}, \href {https://ui.adsabs.harvard.edu/abs/2016PhRvD..94l3004G} {94, 123004}

\bibitem[\protect\citeauthoryear{Gralla \& Lupsasca}{Gralla \& Lupsasca}{2020}]{gralla2020lensing}
Gralla S.~E.,  Lupsasca A.,  2020, Physical Review D, 101, 044031

\bibitem[\protect\citeauthoryear{Gralla, Holz  \& Wald}{Gralla et~al.}{2019}]{gralla2019black}
Gralla S.~E.,  Holz D.~E.,   Wald R.~M.,  2019, Physical Review D, 100, 024018

\bibitem[\protect\citeauthoryear{Gurvits et~al.,}{Gurvits et~al.}{2022}]{gurvits2022science}
Gurvits L.~I.,  et~al., 2022, Acta astronautica, 196, 314

\bibitem[\protect\citeauthoryear{{Guzm{\'a}n}}{{Guzm{\'a}n}}{2006}]{2006PhRvD..73b1501G}
{Guzm{\'a}n} F.~S.,  2006, \mn@doi [\prd] {10.1103/PhysRevD.73.021501}, \href {https://ui.adsabs.harvard.edu/abs/2006PhRvD..73b1501G} {73, 021501}

\bibitem[\protect\citeauthoryear{Johnson et~al.,}{Johnson et~al.}{2020}]{johnson2020universal}
Johnson M.~D.,  et~al., 2020, Science advances, 6, eaaz1310

\bibitem[\protect\citeauthoryear{Johnson et~al.,}{Johnson et~al.}{2023}]{johnson2023key}
Johnson M.~D.,  et~al., 2023, Galaxies, 11, 61

\bibitem[\protect\citeauthoryear{Johnson, Akiyama, Baturin  et~al.}{Johnson et~al.}{2024}]{johnson2024bhex}
Johnson M.~D.,  Akiyama K.,  Baturin R.,   et~al., 2024, arXiv preprint arXiv:2406.12917

\bibitem[\protect\citeauthoryear{Kocherlakota, Rezzolla, Roy  \& Wielgus}{Kocherlakota et~al.}{2024}]{kocherlakota2024hotspots}
Kocherlakota P.,  Rezzolla L.,  Roy R.,   Wielgus M.,  2024, Monthly Notices of the Royal Astronomical Society, p. stae1321

\bibitem[\protect\citeauthoryear{{Krut}, {Arg{\"u}elles}, {Chavanis}, {Rueda}  \& {Ruffini}}{{Krut} et~al.}{2023}]{2023ApJ...945....1K}
{Krut} A.,  {Arg{\"u}elles} C.~R.,  {Chavanis} P.~H.,  {Rueda} J.~A.,   {Ruffini} R.,  2023, \mn@doi [\apj] {10.3847/1538-4357/acb8bd}, \href {https://ui.adsabs.harvard.edu/abs/2023ApJ...945....1K} {945, 1}

\bibitem[\protect\citeauthoryear{{Mazur} \& {Mottola}}{{Mazur} \& {Mottola}}{2004}]{mazur2004}
{Mazur} P.~O.,  {Mottola} E.,  2004, \mn@doi [Proceedings of the National Academy of Science] {10.1073/pnas.0402717101}, \href {https://ui.adsabs.harvard.edu/abs/2004PNAS..101.9545M} {101, 9545}

\bibitem[\protect\citeauthoryear{{McClintock}, {Narayan}  \& {Rybicki}}{{McClintock} et~al.}{2004}]{2004ApJ...615..402M}
{McClintock} J.~E.,  {Narayan} R.,   {Rybicki} G.~B.,  2004, \mn@doi [\apj] {10.1086/424474}, \href {https://ui.adsabs.harvard.edu/abs/2004ApJ...615..402M} {615, 402}

\bibitem[\protect\citeauthoryear{{Mestre}, {Arg{\"u}elles}, {Carpintero}, {Crespi}  \& {Krut}}{{Mestre} et~al.}{2024}]{2024arXiv240419102M}
{Mestre} M.~F.,  {Arg{\"u}elles} C.~R.,  {Carpintero} D.~D.,  {Crespi} V.,   {Krut} A.,  2024, \mn@doi [arXiv e-prints] {10.48550/arXiv.2404.19102}, \href {https://ui.adsabs.harvard.edu/abs/2024arXiv240419102M} {p. arXiv:2404.19102}

\bibitem[\protect\citeauthoryear{{Millauro}, {Arg{\"u}elles}, {Vieyro}, {Crespi}  \& {Mestre}}{{Millauro} et~al.}{2024}]{millauro2024}
{Millauro} C.,  {Arg{\"u}elles} C.~R.,  {Vieyro} F.~L.,  {Crespi} V.,   {Mestre} M.~F.,  2024, \mn@doi [Astronomy \& Astrophysics] {10.1051/0004-6361/202348461}, \href {https://ui.adsabs.harvard.edu/abs/2024A&A...685A..24M} {685, A24}

\bibitem[\protect\citeauthoryear{Mummery, Jiang  \& Fabian}{Mummery et~al.}{2024a}]{mummery2024plunging}
Mummery A.,  Jiang J.,   Fabian A.,  2024a, arXiv preprint arXiv:2406.14957

\bibitem[\protect\citeauthoryear{Mummery, Ingram, Davis  \& Fabian}{Mummery et~al.}{2024b}]{mummery2024continuum}
Mummery A.,  Ingram A.,  Davis S.,   Fabian A.,  2024b, Monthly Notices of the Royal Astronomical Society, 531, 366

\bibitem[\protect\citeauthoryear{{Narayan} \& {McClintock}}{{Narayan} \& {McClintock}}{2008}]{2008NewAR..51..733N}
{Narayan} R.,  {McClintock} J.~E.,  2008, \mn@doi [New Astronomy Reviews] {10.1016/j.newar.2008.03.002}, \href {https://ui.adsabs.harvard.edu/abs/2008NewAR..51..733N} {51, 733}

\bibitem[\protect\citeauthoryear{{Olivares} et~al.,}{{Olivares} et~al.}{2020}]{2020MNRAS.497..521O}
{Olivares} H.,  et~al., 2020, \mn@doi [Monthly Notices of the Royal Astronomical Society] {10.1093/mnras/staa1878}, \href {https://ui.adsabs.harvard.edu/abs/2020MNRAS.497..521O} {497, 521}

\bibitem[\protect\citeauthoryear{Pelle, Reula, Carrasco  \& Bederian}{Pelle et~al.}{2022}]{pelle2022skylight}
Pelle J.,  Reula O.,  Carrasco F.,   Bederian C.,  2022, \mn@doi [Monthly Notices of the Royal Astronomical Society] {10.1093/mnras/stac1857}

\bibitem[\protect\citeauthoryear{{Perlick} \& {Tsupko}}{{Perlick} \& {Tsupko}}{2022}]{2022PhR...947....1P}
{Perlick} V.,  {Tsupko} O.~Y.,  2022, \mn@doi [Physics Reports] {10.1016/j.physrep.2021.10.004}, \href {https://ui.adsabs.harvard.edu/abs/2022PhR...947....1P} {947, 1}

\bibitem[\protect\citeauthoryear{Rosa, Macedo  \& Rubiera-Garcia}{Rosa et~al.}{2023}]{rosa2023imaging}
Rosa J.~L.,  Macedo C.~F.,   Rubiera-Garcia D.,  2023, Physical Review D, 108, 044021

\bibitem[\protect\citeauthoryear{{Ruffini}, {Arg{\"u}elles}  \& {Rueda}}{{Ruffini} et~al.}{2015}]{2015MNRAS.451..622R}
{Ruffini} R.,  {Arg{\"u}elles} C.~R.,   {Rueda} J.~A.,  2015, \mn@doi [Monthly Notices of the Royal Astronomical Society] {10.1093/mnras/stv1016}, \href {http://adsabs.harvard.edu/abs/2015MNRAS.451..622R} {451, 622}

\bibitem[\protect\citeauthoryear{{Saxton}, {Younsi}  \& {Wu}}{{Saxton} et~al.}{2016}]{2016MNRAS.461.4295S}
{Saxton} C.~J.,  {Younsi} Z.,   {Wu} K.,  2016, \mn@doi [Monthly Notices of the Royal Astronomical Society] {10.1093/mnras/stw1626}, \href {https://ui.adsabs.harvard.edu/abs/2016MNRAS.461.4295S} {461, 4295}

\bibitem[\protect\citeauthoryear{Schnittman, Krolik  \& Noble}{Schnittman et~al.}{2016}]{schnittman2016disk}
Schnittman J.~D.,  Krolik J.~H.,   Noble S.~C.,  2016, The Astrophysical Journal, 819, 48

\bibitem[\protect\citeauthoryear{Shakura \& Sunyaev}{Shakura \& Sunyaev}{1973}]{shakura1973}
Shakura N.~I.,  Sunyaev R.~A.,  1973, Astronomy and Astrophysics, Vol. 24, p. 337-355, 24, 337

\bibitem[\protect\citeauthoryear{Tiede, Johnson, Pesce, Palumbo, Chang  \& Galison}{Tiede et~al.}{2022}]{tiede2022measuring}
Tiede P.,  Johnson M.~D.,  Pesce D.~W.,  Palumbo D.~C.,  Chang D.~O.,   Galison P.,  2022, Galaxies, 10, 111

\bibitem[\protect\citeauthoryear{{Vincent}, {Meliani}, {Grandcl{\'e}ment}, {Gourgoulhon}  \& {Straub}}{{Vincent} et~al.}{2016}]{vincent2016}
{Vincent} F.~H.,  {Meliani} Z.,  {Grandcl{\'e}ment} P.,  {Gourgoulhon} E.,   {Straub} O.,  2016, \mn@doi [Classical and Quantum Gravity] {10.1088/0264-9381/33/10/105015}, \href {https://ui.adsabs.harvard.edu/abs/2016CQGra..33j5015V} {33, 105015}

\bibitem[\protect\citeauthoryear{Vincent, Wielgus, Abramowicz, Gourgoulhon, Lasota, Paumard  \& Perrin}{Vincent et~al.}{2021}]{vincent2021}
Vincent F.,  Wielgus M.,  Abramowicz M.,  Gourgoulhon E.,  Lasota J.-P.,  Paumard T.,   Perrin G.,  2021, Astronomy \& Astrophysics, 646, A37

\bibitem[\protect\citeauthoryear{{Visser}}{{Visser}}{2014}]{2014PhRvD..90l7502V}
{Visser} M.,  2014, \mn@doi [\prd] {10.1103/PhysRevD.90.127502}, \href {https://ui.adsabs.harvard.edu/abs/2014PhRvD..90l7502V} {90, 127502}

\bibitem[\protect\citeauthoryear{{Visser} \& {Wiltshire}}{{Visser} \& {Wiltshire}}{2004}]{visser2004}
{Visser} M.,  {Wiltshire} D.~L.,  2004, \mn@doi [Classical and Quantum Gravity] {10.1088/0264-9381/21/4/027}, \href {https://ui.adsabs.harvard.edu/abs/2004CQGra..21.1135V} {21, 1135}

\bibitem[\protect\citeauthoryear{Younsi, Psaltis  \& {\"O}zel}{Younsi et~al.}{2023}]{younsi2023black}
Younsi Z.,  Psaltis D.,   {\"O}zel F.,  2023, The Astrophysical Journal, 942, 47

\bibitem[\protect\citeauthoryear{{Yunis}, {Arg{\"u}elles}, {Mavromatos}, {Molin{\'e}}, {Krut}, {Carinci}, {Rueda}  \& {Ruffini}}{{Yunis} et~al.}{2020}]{2020PDU....3000699Y}
{Yunis} R.,  {Arg{\"u}elles} C.~R.,  {Mavromatos} N.~E.,  {Molin{\'e}} A.,  {Krut} A.,  {Carinci} M.,  {Rueda} J.~A.,   {Ruffini} R.,  2020, \mn@doi [Physics of the Dark Universe] {https://doi.org/10.1016/j.dark.2020.100699}, \href {https://ui.adsabs.harvard.edu/abs/2020PDU....3000699Y} {30, 100699}

\makeatother
\end{thebibliography}




\appendix

\section{Radius of the observation points}
\label{app:distance}

In the following, we analyze the validity of taking observation points at $r=1\,$pc to estimate observable quantities associated with the radiation field at much larger distances. For this, we consider the deviation angles in the Newtonian limit, which we can reasonably apply at radii larger than $1\,$pc. 

A static metric in the Newtonian limit has the form
\begin{equation}
   ds^2 = -(1+2\phi) dt^2 + (1-2\phi) d\gamma^2\,,
   \label{eq:newtonian_metric}
\end{equation}
where $\phi$ is identified as the Newtonian gravitational potential, and $d\gamma^2$ is the three-dimensional flat metric. We can decompose the metric above into a flat background plus a small Newtonian perturbation. In an equal manner, the geodesics of the perturbed spacetime decompose as
\begin{equation}
   x^\mu(\lambda) = x^{(0)\mu}(\lambda) + x^{(1)\mu}(\lambda)\,, 
\end{equation}
where the first term of the right-hand side is the background trajectory, and the second term is a small deviation from the background.
The background trajectory is a geodesic of the background spacetime (a straight line since the background is flat). Besides, we define
\begin{equation}
   k^{\mu} = \frac{d x^{(0)\mu}}{d \lambda} \quad \text{and} \quad l^{\mu} = \frac{d x^{(1)\mu}}{d \lambda}\,, 
\label{eq:tangents}
\end{equation}
where we denote spatial three-vectors in boldface. Since we are interested in null geodesics, $k^\mu$ must satisfy $(k^t)^2 = \bm{k}^2 =: k^2 $, and we can assume $k=1$ without loss of generality. We can derive the following first-order relation for the deviation vector from the geodesic equations \citep{carroll2019spacetime}:
\begin{equation}
   \frac{d \bm{l}}{d \lambda} = -2 [ \bm{\nabla} \phi - \left( \bm{\nabla} \phi \cdot \bm{k} \right) \bm{k}]\,,
   \label{eq:l_vector}
\end{equation}
where evaluation along the background path is implied.

Applying this to our model, in the RAR metric of Eq.~(\ref{eq:rar_metric}), we expand $e^\nu \approx 1 + \nu$ at large radii, where $\nu$ is small, identifying $\phi = \nu / 2$ as the Newtonian potential. We can cast the resulting metric into the form of the Newtonian limit by rescaling the radius as 
\begin{equation}
   \tilde{r} = r \exp \left(-\frac{1}{2}\int_0^r \frac{\nu}{r^\prime} dr^\prime \right)\,.  
   \label{eq:rescaling}
\end{equation}
The quotient $\tilde{r}/r$ remains close to $3\%$ for all $r > 1\,$pc. 

To test the extent to which the DM distribution can influence light at large distances, we integrated Eq.~(\ref{eq:l_vector}) along an ingoing geodesic between $r = 8\,$kpc and $r = 1\,$pc for each RAR solution considered in this work. We take the initial spatial momentum of the geodesic forming an angle $\delta = 50 r_\text{sat} / (8\,\text{kpc})$ with the radial direction, while the initial deviation vector is zero. The norms of the final deviation vectors, shown in Table \ref{tab:deviations}, are negligible in all scenarios. This fact means that any significant deflection of light must occur within $r = 1\,$pc, and we conclude that it is valid to take the observation points at $1\,$pc as we did for our radiative transfer calculations.

\begin{table}
\centering
\sisetup{table-format=1.2e+1} 
\begin{tabular}{c | c c}
\toprule
\quad $m \,$(keV) \quad & \quad $M_{\rm c}\,$($M_{\odot}$) \quad & \quad \(\|\mathbf{l}\|\) \\
\midrule
48  & \num{1e7} & $\num{6.2e-12}$ \\
155 & \num{1e7} & $\num{2.5e-13}$ \\
200 & \num{1e7} & $\num{1.1e-13}$ \\
200 & \num{3.5e6} & $\num{6.8e-14}$ \\
300 & \num{3.5e6} & $\num{2e-14}$ \\
378 & \num{3.5e6} & $\num{6.6e-15}$ \\
\bottomrule
\end{tabular}
\caption{Norm of the final deviation vector for an ingoing geodesic in the Newtonian approximation between $r = 8\,$kpc and $r = 1\,$pc for the RAR solutions we consider in this work.}
\label{tab:deviations}
\end{table}

\bsp	
\label{lastpage}
\end{document}